\documentclass{article}
\usepackage{graphicx, animate}
\usepackage[utf8]{inputenc}
\usepackage{makeidx}
\usepackage{xmpmulti}
\usepackage{hyperref}
\usepackage{natbib}
\usepackage{array, booktabs, tabularx, float, changepage}
\usepackage[table]{xcolor}
\usepackage{amsmath, bm}
\usepackage{soul}
\usepackage{multirow}
\usepackage{epstopdf}
\AppendGraphicsExtensions{.gif}
\include{include/usepackage}
\include{include/preamble}
\include{include/macros}
\usepackage[paperheight=24cm,paperwidth=20cm,textwidth=12cm]{geometry}

\DeclareUnicodeCharacter{03B5}{\ensuremath{\varepsilon}}
\DeclareUnicodeCharacter{2010}{-}
\title{Effects of Molecular Composition and Chain Length on the Interfacial and Thermodynamic Properties of Cyclic and Linear Polymer Blends}
\author{ Oluwatumininu E. Ayo-Ojo$^{a}$\\ 
  \texttt{220052876@stu.ukzn.ac.za}\\
  \texttt{oluwatumininuayoojo@gmail.com}
  \and \\
  Nkosinathi Dlamini$^{a}$ \\ 
  \texttt{dlaminin2@ukzn.ac.za}\\
  \\
  \\
  $^{a}$School of Chemistry \& Physics, University of KwaZulu-Natal,\\
  Pietermaritzburg, Private Bag X01, Scottsville 3209, South Africa}

\begin{document}
\maketitle
\newpage

\section*{Abstract}
Understanding how polymer architecture and composition influence interfacial and thermodynamic properties is critical for advancing sustainable materials. We employed molecular dynamics simulations using the Kremer-Grest bead-spring model to systematically investigate the effects of chain length and blend composition on the adsorption, heat capacity, energy partitioning, and surface tension of cyclic and linear polymer blends at solid interfaces. By varying chain lengths (10, 20, 40, and 60 monomers) and cyclic polymer concentrations (10\% and 90\%), we quantified how these parameters modulate local composition profiles, thermal stability, and interfacial behavior. We found that longer cyclic chains preferentially localize near interfaces, heat capacity exhibits non-linear dependence on chain length and blend ratio, and energy partitioning highlights the dominance of pairwise interactions in short-chain blends. These results provide quantitative insight into the structure–property relationships underpinning polymer blend performance and degradability, informing the design of environmentally friendly polymers with optimized lifecycle characteristics.

\bigskip
\textbf{Keywords:} Polymer blends; Fourier transform ; Infrared Spectroscopy; Molecular dynamics; Chain architecture; Interface; Thermodynamics.

\newpage

\section*{Introduction}
Polymer topology strongly influences the structural and dynamic characteristics of polymer systems\cite{A1}. This has prompted extensive research into how the repetitive chemical units in polymer chains impact interface diffusion. Many studies focus on polymer blends composed of chains with identical chemical compositions but varying molecular architectures, such as linear and cyclic forms. However, a thorough understanding of how molecular weight and chain architecture determine preferential adsorption at interfaces remains limited\cite{A2}.

For linear and cyclic polymer blends, self-consistent field theory predicts that cyclic polymers will accumulate more at interfaces\cite{A3}, regardless of molecular weight, but experimental observations have contradicted this expectation\cite{A4}. Electron spectroscopy for chemical analysis (ESCA) on blends of linear and cyclic polystyrene revealed that linear chains enrich the surface when cyclic chains are present at low concentrations\cite{A5}. This highlights inconsistencies for non-freely jointed chain polymer conformations, especially in shorter chains.

The environmental impact of synthetic polymers is a growing concern due to their contribution to plastic pollution. While not all polymers are biodegradable, recent research has highlighted the potential for enzymatic degradation of certain polymers\cite{A10,A11,A12}. Identifying and designing enzymes that break down polymers into reusable constituents is a promising step toward sustainability. Ideally, this approach would enable polymers to degrade into their original building blocks, allowing for their reuse in synthesizing new materials and achieving a truly circular lifecycle\cite{A13}. Achieving this goal requires a deeper understanding of the interfacial and stress-related properties of polymers with varying structures. These properties are key to optimizing polymer blends for performance and degradability, providing insight into how polymers, with regard to their architecture, interact with surfaces and respond to environmental stressors.

One area requiring further exploration is the role of polymer structure in determining specific heat capacity values. Established methods address the heat capacities of well-defined polymer structures, but complexities arise when modular architectures like block copolymers or amphiphilic polymers are considered. Further understanding these mechanisms could refine predictions of heat capacity in more complex polymer systems, enabling improved materials design.

Hydration significantly affects the heat capacity properties of polymers, particularly in hydrophilic systems. For example, hydration environments can modify the heat capacity of amphiphilic polytartaramides, underscoring the role of water in complex mixtures\cite{A42}. This observation prompts further investigation into how varying polymeric fluid quantities might influence heat capacity, overall thermal stability, and the structural integrity of polymer matrices.

While considerable progress has been made in exploring the relationship between polymers and heat capacity, notable research gaps persist. The central focus of this study is to understand how polymer structure influences heat capacity and to explain the mechanisms behind this structural effect. Heat capacity plays a key role in thermal degradation processes. For instance, polymers with specific thermal properties can undergo photo-induced degradation effectively when exposed to UV light, as the absorbed thermal energy can accelerate the breakdown of polymer chains\cite{A41}. Materials with lower heat capacities might degrade faster than those with higher capacities, as the thermal energy required to reach degradation thresholds is more readily achieved in low-capacity polymers.

A high energy partitioning ratio, indicative of dominant pairwise interaction energies over bonded interaction energies, suggests a less cohesive polymer structure\cite{A45}. This lack of strong inter-chain interactions may predispose the polymer to degradation, as weaker connections can be broken more easily compared to polymers with more robust linked networks. This dynamic allows for a higher likelihood of enzymatic or reactive attack on the polymer chains, as indicated by studies on the behavior of polymers under varying thermal and chemical conditions.

Thermal degradation of polymer blends, such as polyvinyl alcohol (PVA), is associated with local molecular motions reflected in their heat capacity\cite{A43}. This thermodynamic behavior can enhance polymer degradability, as more mobile molecular chains may facilitate chemical interactions such as hydrolysis or enzymatic degradation that contribute to breaking down the polymer.

Degradation processes are often dictated by the polymer's structure and its interactions at the molecular level. For example, the degradation of polymers typically occurs due to the breaking of long molecular chains, which directly links to properties such as viscosity, also influenced by heat capacity\cite{A44}. Thus, materials characterized by higher mobility and lower bonding interactions, as indicated by lower energy ratios in energy partitioning plots, should experience degradation more readily because the necessary mechanical and chemical bonds needed for stability are less robust.

In this study, we explore the behaviors of polymer blends, focusing on how molecular weight and chain architecture—particularly linear and cyclic topologies—govern preferential adsorption near solid interfaces. By examining these factors, we enhance understanding of how polymer structure influences key thermal, physical, and stress-related properties. This insight is crucial for clarifying the relationship between polymer architecture and degradability, supporting the development of environmentally sustainable materials with improved lifecycle performance.

\subsection*{Models}
We used the Kremer–Grest bead-spring model\cite{A14} to investigate the behavior of cyclic and linear polymers at the polymer/vacuum interface. This model is well suited for studying various polymeric systems. Polymers are represented as chains of sequential monomers with uniform mass (\(m\)), where each monomer connects to its neighbors to form either cyclic (closed-loop) or linear (open-loop) configurations.

The interaction potential for non-bonded monomers is described by the truncated and shifted Lennard-Jones (LJ) potential\cite{A15} at \(r_c) = 2.5\sigma\):
\begin{equation}
E_{\text{LJ}}(r) =
\begin{cases} 
4\epsilon \left[ \left( \frac{\sigma}{r} \right)^{12} - \left( \frac{\sigma}{r} \right)^6 \right] + \epsilon_{\text{LJ}}, & r \leq r_c \\ 
0, & r > r_c,
\end{cases}
\end{equation}
where \( r \) is the distance between two monomers, \( \epsilon \) is the depth of the potential well, and \( \epsilon_{\text{LJ}} \) is a constant ensuring continuity of the potential at \( r = r_c \).

Bonded monomers interact via the LJ potential, coupled with the finitely extensible nonlinear elastic (FENE) potential\cite{A16}:
\begin{equation}
E_f(r) =
\begin{cases} 
-\frac{1}{2} K r_0^2 \ln \left[ 1 - \left( \frac{r}{r_0} \right)^2 \right], & r \leq r_0 \\ 
\infty, & r > r_0,
\end{cases}
\end{equation}
where $r_0 = 1.5\sigma$ is the maximum bond extension, and $K = 30\epsilon/\sigma^2$. Overlap between three consecutive beads is prevented, and monomers experience both Brownian and frictional forces\cite{A17}.

\subsection*{Simulation Details}
We generated initial configurations by randomly placing cyclic and linear polymers of equal chain lengths within a parallelepiped simulation box. The study considered polymer chains of 10, 20, 40, and 60 monomers. The concentration of cyclic polymers in the blend ($C_0$) was set to 10\% (C10) and 90\% (C90) for all chain lengths, calculated as:
\begin{equation}
C_0 = 100 \times \frac{N_c^t}{N_c^t + N_l^t},
\end{equation}
where $N_c^t$ and $N_l^t$ represent the total number of cyclic and linear polymer monomers, respectively. Each blend contained 120,000 monomers.

We performed simulations using the LAMMPS package\cite{A18}, employing a velocity-Verlet algorithm with a timestep of $\Delta t = 0.005\tau$, where $\tau = (m\sigma^2/\epsilon)^{1/2}$. The systems were equilibrated in the NPT ensemble at $P = 0$ and $T = \epsilon/k_B$, using a Langevin thermostat and a Berendsen barostat, with periodic boundary conditions along all three directions. After equilibration, we removed periodicity along the $z$-axis and elongated the simulation box in this direction to expose the polymers to nearly equal volumes of a wall interface at both the top and bottom. This allowed us to evaluate the influence of interfacial confinement on the polymers. Simulations were conducted in the presence of walls to enable direct interaction with such interfaces, providing a more accurate analysis of polymer behavior under realistic confinement conditions.

\section*{Results and Discussion}
\subsection*{Local Composition}
We calculated the local composition as the relative proportion of each polymer type in the blend. The composition of cyclic and linear polymers is given by:
\begin{equation}
    \text{Composition of Cyclic Polymers (\%)} = \frac{N_\text{cyclic}}{N_\text{total}} 
\end{equation}
\begin{equation}
    \text{Composition of Linear Polymers (\%)} = \frac{N_\text{linear}}{N_\text{total}} 
\end{equation}
where $ N_\text{cyclic} $ and $ N_\text{linear} $ are the number of monomers belonging to cyclic and linear polymers, respectively, and $ N_\text{total} $ is the total number of monomers in the system.

We implemented this calculation using a Fortran code to parse the LAMMPS output dump file for the numbers of cyclic and linear monomers at the varying compositions, using slabs of size $\sigma$ along the $z$-direction.

\subsubsection*{Cyclic Polymers at 10\% Composition}
\begin{figure}[h!]
    \centering
    \includegraphics[width=0.6\textwidth]{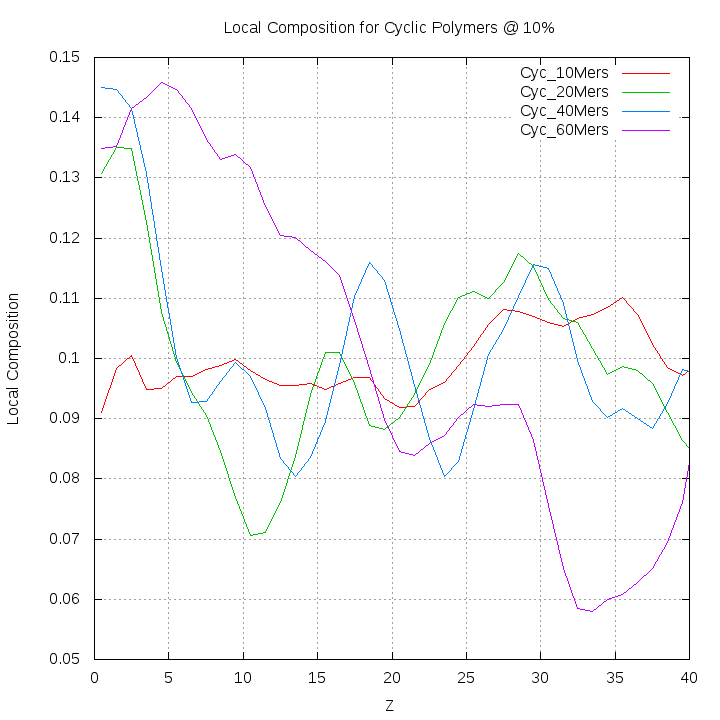}
    \caption{Enrichment of longer cyclic polymer chains at the wall interface in blends with 10\% cyclic content. Local composition profiles are shown for cyclic polymers of varying chain lengths (10, 20, 40, and 60 monomers) as a function of distance from the wall ($Z=0$). Longer cyclic chains preferentially localize near the interface, while shorter chains are depleted.}
    \label{fig:cyc10}
\end{figure}

Figure \ref{fig:cyc10} shows that the concentration of short polymer chains (10-mers) decreases near the wall ($Z \approx 0$), while longer chains (over 20-mers) tend to localize and enhance their concentration in that region. This behavior results from entropic penalties faced by cyclic polymers, which tend to adopt compact configurations that limit their interactions with the wall\cite{A21,A46}. Longer cyclic polymers have more degrees of freedom but incur significant entropic costs when constrained in bulk. When these longer polymers approach a wall, they localize to minimize entropic penalty, leading to greater concentration near the interface\cite{A21}. Polymer-wall interactions can draw polymers into the interface, leading to alignment parallel to the wall\cite{A46}. Attractive interactions between the polymer chains and the wall promote this aggregation effect, increasing their concentration near the wall.

\subsubsection*{Cyclic Polymers at 90\% Composition}
\begin{figure}[H]
    \centering
    \includegraphics[width=0.6\textwidth]{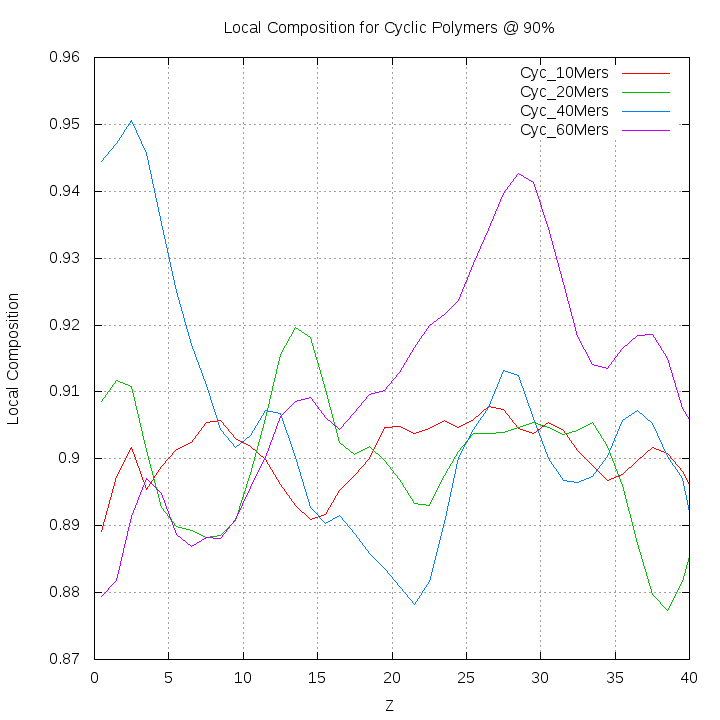}
    \caption{Influence of chain length on interfacial localization in blends with 90\% cyclic content. Local composition profiles for cyclic polymers of different chain lengths (10, 20, 40, and 60 monomers) show enhanced accumulation of longer chains near the wall interface ($Z=0$), with only slight depletion in the bulk region ($Z>20$).}
    \label{fig:cyc90}
\end{figure}
At lower concentrations of cyclic polymers, where interactions are weaker due to compact architecture, the differentiation in behaviour between short and long chains is more pronounced. As polymer chain lengths increase, the tendency for these polymers to intercalate and form stable arrangements near the wall becomes stronger due to their spatial configurations and interactions\cite{A48}. In Figure \ref{fig:cyc90}, the bulk region ($Z > 20$) experiences only slight depletion, influenced by stronger enhancement near the wall, affirming the dominance of chain length on the spatial distribution of polymers.

The overall density of polymer chains near the wall is influenced by both length and crowding effects at higher concentrations. Enthalpic interactions further diminish entropic drives, resulting in significant accumulations near the wall\cite{A21}. Longer cyclic polymers are adept at forming stable conformations in this region, making them more likely to enhance the local composition in concentrated blends. These trends are consistent with studies on chain length and concentration in relation to wall interactions\cite{A46,A47}.

\begin{figure}[H]
    \centering
    \includegraphics[width=0.38\linewidth]{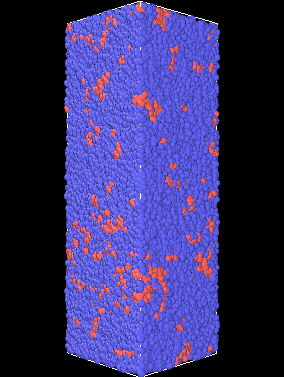}
    \caption{Simulation snapshot illustrating spatial distribution of polymers in a blend with 90\% cyclic and 10\% linear chains (40 monomers each). Red and blue colors represent cyclic and linear polymers, respectively. Both polymer types are observed near the walls, confirming the composition profiles.}
    \label{fig:MD_sim}
\end{figure}

Figure \ref{fig:MD_sim} shows a simulation snapshot for a polymer blend comprising 40-mers, with 10\% cyclic and 90\% linear polymers. This confirms the presence of both polymer types near the walls, supporting the local composition plots.

\subsection*{Temperature and Total Energy Calculation}
The temperature in molecular dynamics (MD) represents the kinetic energy of the atoms or molecules in the system\cite{A25}:
\begin{equation}
T = \frac{2K}{3N \cdot k_B}
\end{equation}
where $K$ is total kinetic energy, $N$ is number of degrees of freedom, and $k_B$ is Boltzmann constant.

The total energy is the sum of kinetic and potential energy:
\begin{equation}
E_{\text{total}} = K + U
\end{equation}

\subsubsection*{Heat Capacity Calculation}
We calculated heat capacity at constant volume ($C_v$) using the fluctuation formula from statistical mechanics\cite{A26}:
\begin{equation}
C_v = \frac{\langle E^2 \rangle - \langle E \rangle^2}{k_B T^2}
\end{equation}
where $E$ is total energy, $\langle E \rangle$ is average total energy, $\langle E^2 \rangle$ is mean of the square of total energy, $k_B$ is Boltzmann constant, and $T$ is average temperature.

\begin{figure}[H]
    \centering

    \begin{minipage}[b]{0.48\linewidth}
        \centering
        \includegraphics[width=\linewidth]{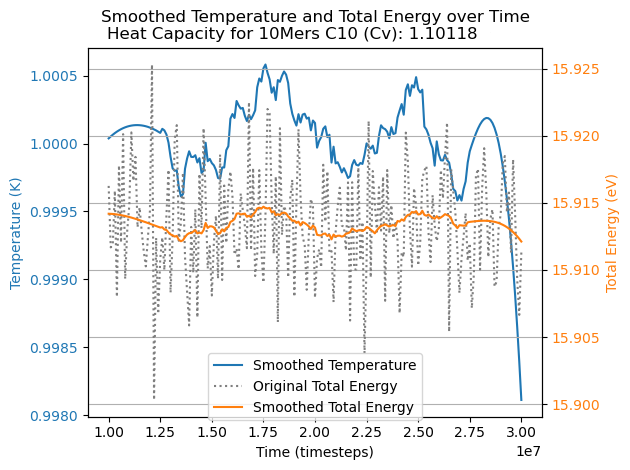}
        \textbf{(a) 10Mers Polymers}
    \end{minipage}
    \hfill
    \begin{minipage}[b]{0.48\linewidth}
        \centering
        \includegraphics[width=\linewidth]{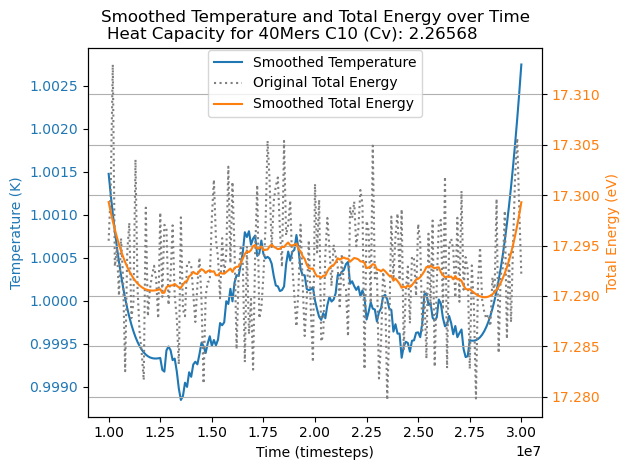}
        \textbf{(b) 40Mers Polymers}
    \end{minipage}

    \caption{Time evolution of temperature and total energy for 10-mer and 40-mer polymer blends with 10\% cyclic content. Smoothed temperature (blue) and total energy (orange) curves demonstrate system equilibration and energy conservation during the calculation of heat capacity ($C_v$).}
    \label{fig:Temp_Energy}
\end{figure}

\begin{figure}[H]
    \centering
  \begin{minipage}[b]{0.48\linewidth}
        \centering
        \includegraphics[width=\linewidth]{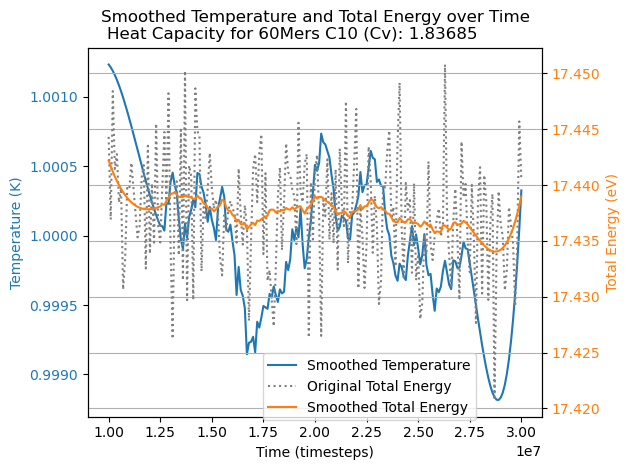}
        \textbf{(a) 60Mers Polymers 10\% Cyclic \& 90\% Linear}
    \end{minipage}
    \hfill
    \begin{minipage}[b]{0.48\linewidth}
        \centering
        \includegraphics[width=\linewidth]{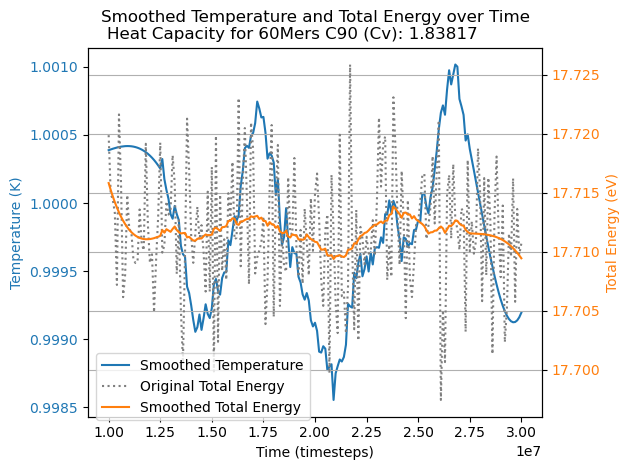}
        \textbf{(b) 60Mers Polymers 90\% Cyclic \& 10\% Linear}
    \end{minipage}
     60Mers Polymers\\
    \caption{Temperature and total energy trajectories for 60-mer polymer blends with 10\% and 90\% cyclic content.
    The plots illustrate thermal stability and energy fluctuations used in heat capacity ($C_v$) calculations for different blend compositions.}
    \label{fig:Temp_Energy+}
\end{figure}

The smoothed temperature curves in Figures \ref{fig:Temp_Energy} and \ref{fig:Temp_Energy+} show fluctuations around a mean value, indicating how well the polymer system maintains thermal stability over time. The total energy trends reflect the stability of the energy states of the system, helping to identify thermal equilibrium. The total energy (orange line) appears relatively stable over time, indicating equilibration and energy conservation. Small oscillations in total energy show the interplay between kinetic and potential energy during the simulation.

\begin{figure}[h!]
    \centering
    \includegraphics[width=0.8\linewidth]{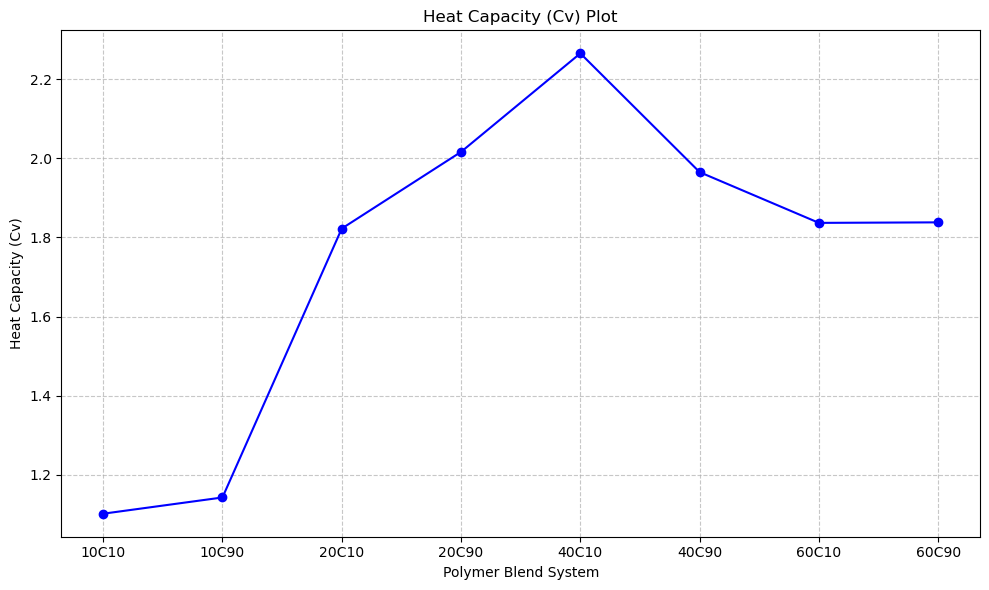}
    \caption{Heat capacity ($C_v$) as a function of chain length and blend composition. Comparison of $C_v$ values for blends with varying proportions of cyclic and linear polymers (10, 20, 40, and 60 monomers) reveals non-linear dependence on chain length and composition.}
    \label{fig:Cv}
\end{figure}

The 10-mer C10 polymer blend demonstrates a higher heat capacity ($C_v$) than its C90 counterpart, as supported by previous literature\cite{A28}. This suggests that C10 systems can absorb more heat energy before a discernible temperature change occurs, implying enhanced thermal responsiveness. The higher heat capacity for systems with a higher proportion of cyclic polymers, particularly in the C90 configuration, can be rationalized by the compact nature of cyclic structures, which exhibit limited configurational degrees of freedom. These characteristics enhance thermal stability, especially in shorter chain configurations\cite{A23}. In longer chains, dynamics may become increasingly restricted, resulting in varied thermal properties.

Polymers with 40 monomer units (40-mers) of C10 exhibit competitive heat capacities among the systems examined. These polymers may facilitate greater thermal energy absorption due to enhanced intra-chain interactions or augmented configurational freedom\cite{A49,A50}. For 60-mer polymers, we observed a reduction in $C_v$ for the C10 blend relative to C90. This may indicate that once the polymer chain length exceeds a threshold, the dynamic behavior of the chains is constrained, leading to lower heat capacities. This pattern emphasizes the relationship between chain length and heat capacity, suggesting a complex interplay rather than a strictly linear correlation\cite{A51}.

\subsection*{Energy Partitioning in the Polymer System}
\begin{figure}[H]
    \includegraphics[width=0.6\linewidth]{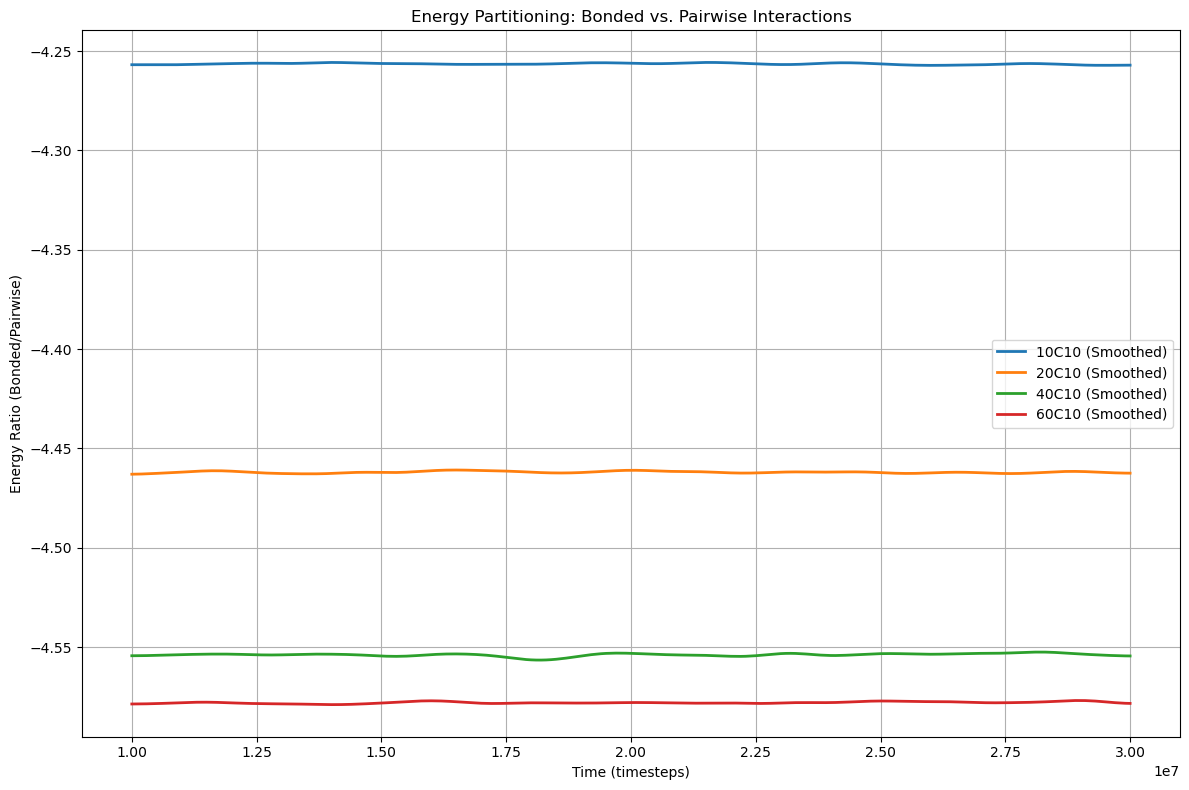} \\
    \includegraphics[width=0.6\linewidth]{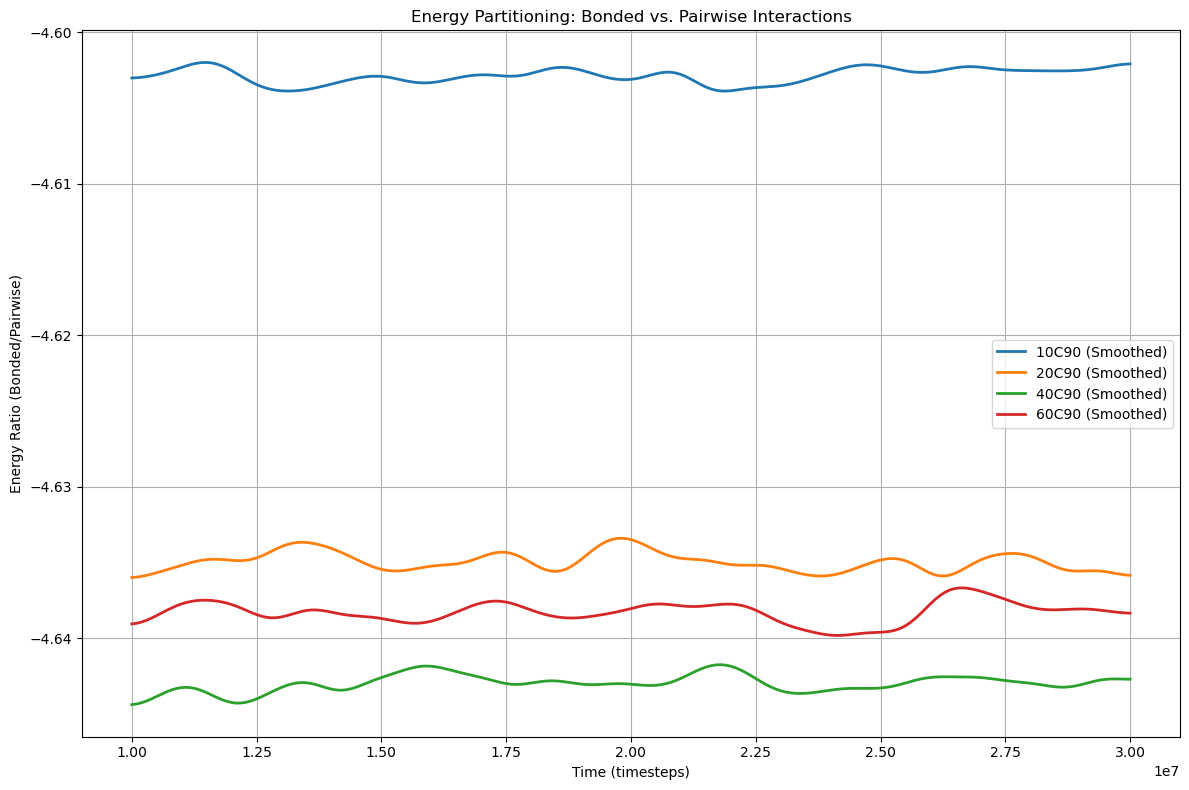}
    \caption{Partitioning of total energy into bonded and pairwise interaction components for polymer blends.
    Energy ratios are shown for blends with different cyclic/linear compositions and chain lengths, highlighting the dominance of pairwise interactions in short-chain blends and the effect of architecture on structural cohesion.}
    \label{fig:EP-BPI}
\end{figure}
Energy partitioning in polymer blends, particularly those with varying ratios of cyclic and linear polymers, reveals distinct behaviors governed by chain architecture. Figure \ref{fig:EP-BPI} shows that the energies remain relatively constant, indicating that the system reached a steady state. The energy ratios (bonded/pairwise interactions) are consistently negative, signifying that pairwise interaction energies dominate over bonded interaction energies in all cases. A high energy ratio would indicate that bonded interactions dominate, suggesting strong structural integrity or chain connectivity\cite{A32}. Blends with 10\% cyclic and 90\% linear content (C10) tend to exhibit higher total energy partitioning than those with 90\% cyclic and 10\% linear content (C90). This arises from the structural properties of cyclic polymers, whose closed-loop conformations can facilitate enhanced interactions with neighboring chains. In contrast, linear polymers predominantly influence pairwise interaction energy due to their greater conformational flexibility\cite{A54}.

As chain length increases, particularly in intermediate regimes such as 20-mers and 40-mers, energy ratios may trend toward more negative values. This could reflect a rise in potential interaction sites that promote intensified pairwise interactions. Cyclic polymers typically exhibit a higher density of contacts, which may encourage greater entanglement and interaction frequency. However, direct empirical data linking chain length to energy partitioning in cyclic and linear polymer systems remains limited, and further investigation is needed\cite{A55}.

\begin{figure}[h!]
    \centering
    \begin{minipage}[b]{0.45\textwidth}
        \centering
        \includegraphics[width=\textwidth]{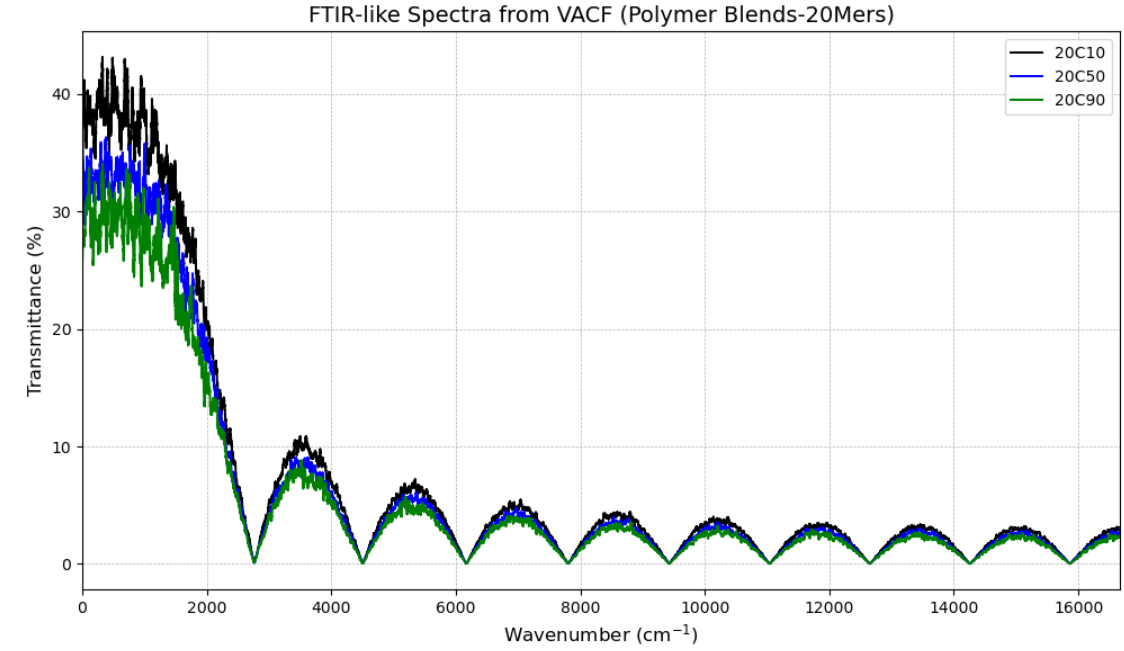}
        \textbf{(a) 20Mers}
    \end{minipage}
    \hfill
    \begin{minipage}[b]{0.45\textwidth}
        \centering
        \includegraphics[width=\textwidth]{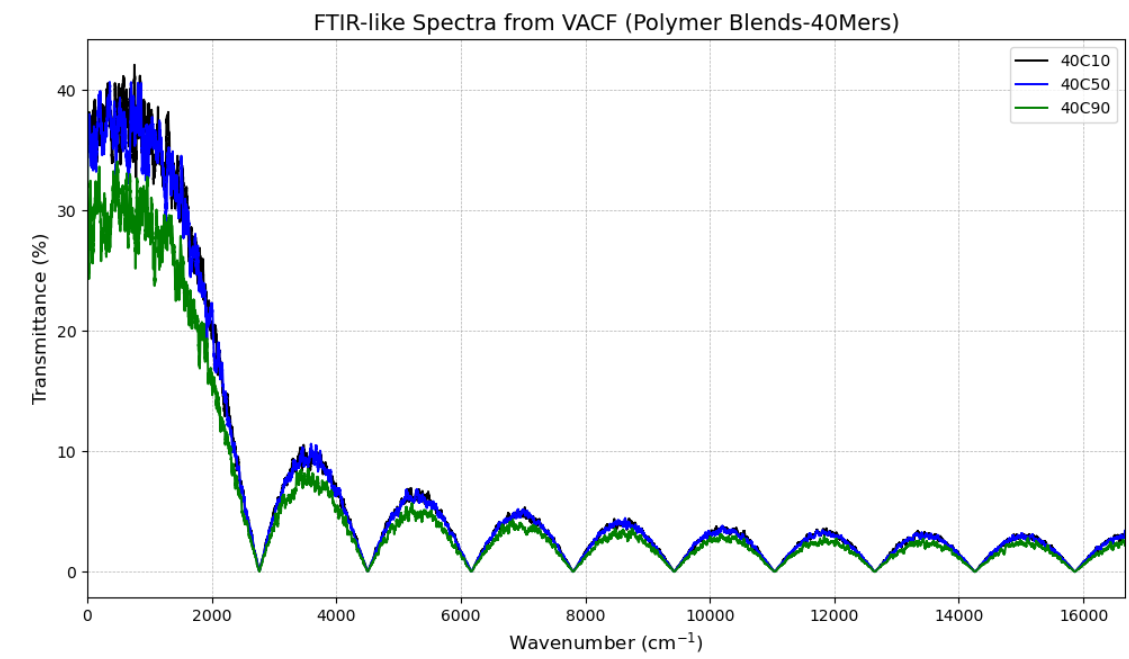}
        \textbf{(b) 40Mers}
    \end{minipage}

    \vspace{0.5cm} 
    \begin{minipage}[b]{0.45\textwidth}
        \centering
        \includegraphics[width=\textwidth]{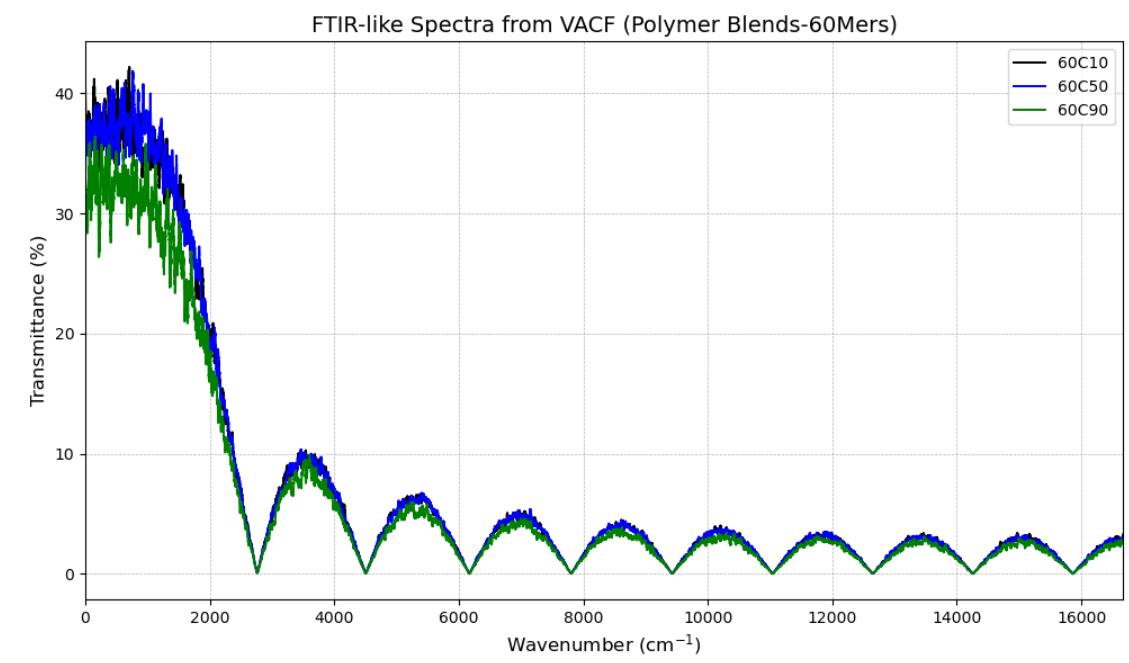}
        \textbf{(c) 60Mers}
    \end{minipage}

    \caption{Fourier Transform Infrared Spectroscopy for polymer blends with varying cyclic content and chain lengths under $z$-axis confinement. 
    The FTIR-like spectra derived from the velocity autocorrelation function (VACF) of cyclic--linear polymer blends provide compelling insights into the vibrational landscape of these systems as a function of chain architecture and strand length. The transmittance spectra for 20Mers (a), 40Mers (b), and 60Mers (c), each at varying cyclic content (10\%, 50\%, 90\%), reveal trends that highlight the interplay between topological constraints, segmental mobility, and vibrational freedom.}
    \label{fig:ftir_spectra}
\end{figure}

\subsection*{Fourier Transform Infrared Spectroscopy}
To compute a theoretical Fourier Transform Infrared (FTIR) spectrum from molecular dynamics simulations, we begin with the \textbf{Velocity Autocorrelation Function (VACF)}, which quantifies how particle velocities are correlated with themselves over time. Using \textsc{LAMMPS}, the VACF can be computed with the \texttt{compute vacf} command for specific polymer groups (e.g., cyclic, linear), and time-averaged using \texttt{fix ave/time}. This gives us VACF data in the time domain.

To convert the VACF into an FTIR-like spectrum, we leverage the fact that the \textbf{Fourier Transform of the VACF yields the Vibrational Density of States (VDOS)}. This spectrum reflects the system's intrinsic vibrational modes. In \textsc{Python}, we apply a Fast Fourier Transform (FFT) to the VACF:

\begin{verbatim}
spectrum = abs(np.fft.fft(vacf_data))
\end{verbatim}

The resulting spectrum is proportional to absorbance, as it captures how vibrational energy is distributed across frequencies. To express this in transmittance (\%), as is standard in FTIR plots, we normalize the spectrum such that its maximum corresponds to 100\%:

\begin{verbatim}
transmittance = (spectrum / np.max(spectrum)) * 100
\end{verbatim}

The \textit{x}-axis of the FTIR spectrum is the \textbf{wavenumber} (cm$^{-1}$), which we compute from the FFT-generated frequency values using the relation:

\[
\tilde{\nu} = \frac{f}{c} \times 10^{-2}
\]

where \( f \) is the frequency in Hz, and \( c \) is the speed of light (approximately \( 3 \times 10^{10} \) cm/s). In \textsc{Python}, this is implemented as:

\begin{verbatim}
frequencies = np.fft.fftfreq(len(vacf_data), d=timestep_in_seconds)
wavenumbers = frequencies / (3.0e10) * 1e-2
\end{verbatim}

This process bridges microscopic velocity dynamics with macroscopic infrared response, allowing us to interpret polymer vibrational behavior just as one would in experimental FTIR spectroscopy.

The analysis of Fourier Transform Infrared Spectroscopy (FTIR) for polymer blends consisting of cyclic and linear polymers of varying lengths and compositions demonstrates specific spectral characteristics intrinsic to the molecular structure and dynamics of the materials studied. In examining the wavenumber axis across all three plots (20Mers, 40Mers, 60Mers) depicted in Figure \ref{fig:ftir_spectra}, it remains unchanged, suggesting that the fundamental vibrational modes are conserved regardless of chain length or cyclic-linear composition. This observation aligns with the understanding that molecular vibrations are dictated predominantly by intramolecular force constants such as bond stiffness, which are less influenced by the overall molecular architecture \cite{A59}.

The FTIR spectra reveal a distinct broad band in the low-wavenumber region ($<$2000~cm$^{-1}$), attributed to collective and segmental motions, including torsions and conformational changes in the backbone of the polymers~\cite{A60}. Furthermore, we observe high-frequency periodic dips ($\sim$3000–1600~cm$^{-1}$), indicating bond stretching vibrations that reflect quantized internal vibrational modes typical of polymer chains~\cite{A61}. A notable decrease in transmittance intensity correlates with increasing cyclic content across all strand lengths. For instance, the 20Mers display stark variations in transmittance, with 20C10 showing values as high as $\sim$42\%, in contrast to the significantly lower transmittance of 20C90. This significant drop in transmittance for the highly cyclic system suggests enhanced suppression of the collective vibrations~\cite{A62, A63}.

While trends persist through the 40Mers and 60Mers samples, an observable flattening in transmittance differences particularly beyond 4000~cm$^{-1}$ indicates that increasing chain length mitigates the vibrational impacts due to topology. This suggests that longer chains experience bulk dynamics which overshadow topological constraints, providing deeper insights into the significance of polymer architecture~\cite{A63}. The consistently high transmittance of low-cyclic-content blends underscores the increased mobility of linear chains, which diminishes as cyclic content escalates due to constrained internal degrees of freedom.

Moreover, the interpretation of these spectra suggests that short linear chains exhibit the highest transmittance, attributed to their flexible nature. Conversely, heightened cyclic content induces localized vibrational characteristics that constrain chain motion, ultimately affecting spectral behavior. In longer chains, such as those represented by the 60Mers, nearly overlapping spectra across varying compositions imply that as chain lengths increase, the effects of topology on vibrational modes begin to average out. This finding implies a critical transition where dynamic behaviors become more universal, primarily influenced by length rather than cyclic or linear architecture~\cite{A64}.

Integrating these findings with conclusions regarding heat capacity and polymer degradation patterns enhances the depth of the analysis. Shorter polymers, particularly oligomers, demonstrate low heat capacity, indicating potential accelerated degradation—a relationship noted to be nonlinear. The data suggest that beyond a molecular weight threshold of approximately 40 monomers, heat capacity diminishes further, indicative of a more significant structural compactness and less cohesive nature that increases susceptibility to degradation~\cite{A63}. Therefore, insights gleaned from our FTIR analysis elucidate vibrational properties but also propose that design considerations for degradable polymers should focus on shorter chain lengths, as these attributes intersect with the dynamics observed in thermal properties.

\section*{Conclusion}
This study demonstrates that the molecular weight and chain architecture of polymers, specifically linear and cyclic topologies, play a crucial role in their interactions at solid interfaces. Although self-consistent field theory predicts that cyclic polymers will gather more readily at these interfaces, experimental data show that linear chains may also enhance surface concentration at lower concentrations. Shorter polymers, such as oligomers, exhibit low heat capacity, which could suggest a propensity for accelerated degradation; however, this relationship is not linear. Once a chain length surpasses approximately 40 monomers, the heat capacity begins to decrease, indicating a threshold effect.

Additionally, shorter polymer chains tend to exhibit the highest energy partitioning, reflecting a dominance of pairwise interactions over bonded interactions. This characteristic implies a less cohesive structure, suggesting that polymers of shorter chains may be more conducive to degradation. The design of degradable polymers would benefit from incorporating shorter chain lengths.

Future research should explore the integration of nano-fillers to modulate the thermal properties of polymers and provide further opportunities for analysis and optimization of polymer performance.

\newpage

\subsection*{Acknowledgments}
We thank the Centre for High-Performance Computing (CHPC) at CSIR for the generous allocation of computational resources.

\subsection*{Data Availability}
The data supporting this study's findings are available from the corresponding author upon reasonable request.

\subsection*{Generative Artificial Intelligence Declaration}
Generative AI tools (ChatGPT, Grammarly) were used to enhance clarity and readability.

\subsection*{Disclosure of Interest}
No potential conflict of interest was reported by the authors.

\subsection*{Funding}
This research was partly funded by the Erasmus+ Mobility Scholarship through Agreement n. 2022-1-IT02\_KA171-HED-000077971.

\section*{Author Contributions}
Conceptualization, investigation, writing—original draft, and data curation: Oluwatumininu E. Ayo-Ojo. Writing—review and editing for intellectual content, and visualization: Oluwatumininu E. Ayo-Ojo and Nkosinathi Dlamini. Supervision and project validation: Nkosinathi Dlamini. All authors have read and agreed to the published version of the manuscript and accept responsibility for all aspects of the work.

\newpage
\bibliography{mybib.bib}
\end{document}